# RF Bucket Reduction as an Experimental Tool for Beam Diagnostics and Manipulation

Xi Yang and C. Ankenbrandt

*Fermi National Accelerator Laboratory*

Box 500, Batavia IL 60510

**Abstract**

A technique using RF bucket reduction for acquiring information about the particle distribution in longitudinal phase space has been applied in the Fermilab Booster. Data sets were obtained at six important time intervals of a Booster cycle for three different beam intensities. Controlled RF bucket reduction also provides other opportunities for beam manipulation.

## 1 Introduction

The 8 GeV beam from the Booster is sent to the Main Injector for further acceleration. The longitudinal emittance (LE) control of the beam in the Booster is especially important for some particular applications, such as slip stacking, proton coalescing, etc. We focus on understanding the relationship between the RFSUM curve during the cycle and the longitudinal emittance growth of the beam and on optimizing the RFSUM curve to make the LE growth a minimum while providing the required amount of acceleration. The various important benefits from the minimum LE growth, such as reducing the transition loss and accelerating higher intensity beam through the Booster cycle, motivate us to develop a technique to acquire information about the particle distribution in longitudinal phase space. Whenever a change is made to the RFSUM curve, the longitudinal phase space distribution and the LE of the beam bunches should be checked.

## 2 Method

Booster RF stations are divided into two groups called A and B. The rf drive to the two groups can be adjusted as a function of time to any relative phase. The A and B stations are balanced so well that the vector sum of the RF voltages from all the cavities in the ring (RFSUM) is close to zero when group A and B stations are electrically 180 degrees out of phase. A low level system curve, which is called the primary paraphase curve, ordinarily controls the phase difference between group A and B stations during rf capture of the beam. The primary paraphase curve brings the voltage vectors of groups A and B into phase relatively slowly such that efficient capture is achieved at injection. By adding a secondary paraphase curve, we can adjust the value of RFSUM at any time in the Booster cycle over the range from zero to a maximum determined by the anode and CIG curves.

This provides a means to achieve RF bucket reduction by reducing RFSUM at a particular time interval in the Booster cycle. During the period of the RF bucket reduction, one should observe no particle loss before the bucket boundary reaches the bunch boundary in longitudinal phase space. We do start to see charge loss when the calculated bucket boundary reaches a plausible value for the bunch LE. Further reduction of the RF bucket will cause more charge loss. Since the RF bucket area can be calculated from the values of RFSUM and synchronous phase, one can obtain a rough determination of the beam emittance distribution from the intensity left in the calculated RF bucket.

An ambiguity exists in the above measurement. One can reduce the RFSUM either slowly (quasi-adiabatically) or rapidly at the beginning of the time interval. Here, the limiting factor for the fast reduction of RFSUM is the Q factor of the RF cavities. The Q factor is several hundred for the Booster, implying that cavity phases can change as fast as several Booster turns. We chose fast reduction of RFSUM to simplify the bucket-area calculation and to avoid most of the complications of time-dependent parameters. For example, a feedback loop works to counter the change made to RFSUM by modifying the synchronous phase. If the synchronous phase is sufficiently changed, the change should be included in the bucket-area calculation. In the reported measurements, we could neglect the feedback system since the particles outside the bucket were lost well before the synchronous phase was sufficiently changed because of the gain limitation in the Booster feedback system. To make the feedback system stable, the gain of the feedback loop is set such that significant change of the synchronous phase takes considerably longer than the fast voltage reduction. Because the reduction of the RF bucket was fast, the initial value of the synchronous phase was used in calculating the bucket area.

## 3 Experimental Result

The most important time intervals in a Booster cycle are right after injection, 5 ms and 10 ms after injection when instabilities start to develop, before and after transition, and before extraction. So we chose six points, 0.9 ms, 5.5 ms, 10.4 ms, 14.4 ms, 19.1 ms, and 29.5 ms in a Booster cycle to do the measurement. We applied the same set of measurements to three different beam intensities. They are two-turn injection, eleven-turn injection and twelve-turn injection. The results are shown in Figs. 1-3.

## 4 Experimental Complication

The way used to reduce the RF bucket area is directly related to the accuracy of the particle distribution measurement in the longitudinal phase space. One concern is how long RFSUM should be kept at its minimum value in the fast reduction case. If RFSUM stays low for too long, the beam will get lost due to a different reason. The insufficient acceleration makes the beam develop a momentum offset to the synchronous particle, until this momentum offset is large enough to reach the dynamical aperture. The beam loss due to off-momentum scraping contributes to the experimental error. See Fig.4 for details. Another concern is the recapture of some particles outside the bucket. If the time taken for particles outside the reduced bucket to be lost is comparably longer than the time taken for RFSUM to recover, some particles will be recaptured. As shown in Fig. 5, there were some particles left even after RFSUM was reduced to zero.

## 5 Application

First, this technique can be applied to experimentally map the boundaries of the particle distribution in longitudinal action or emittance. That can be used to specify an initial RFSUM curve. Afterwards, tuning the initial RFSUM curve based upon the improvement of the transmission efficiency and the reduction of the LE will be used to obtain the optimal RFSUM curve. Since the optimal RFSUM curve is dependent upon the beam intensity, this process should be applied separately for different beam intensities.

At the present running conditions, RFSUM from 17 stations is sufficient for the acceleration. The need for 18 stations is mainly to provide redundancy with one-station failure. The additional accelerating voltage when all cavities are running increases the RF bucket size and contributes to the extra LE. By running different secondary paraphase curves for 17 stations and 18 stations, we can achieve the same optimal RFSUM curve for 17 stations and 18 stations.

By using the RF bucket reduction technique, we can control the time and the amount of the charge to be lost. The benefit of this is to reduce the activation of accelerator components by disposing early in the cycle of protons that are destined to be lost anyway.

## 6 Conclusion

There are some approximations in the present experiment, such as assuming no significant synchronous phase change during the fast bucket reduction and no significant beam loss due to scraping. In the future, we should record the synchronous phase simultaneously when we are reducing the RF bucket area. Besides that, RFSUM should be reduced in such a way that the scraping because of the insufficient acceleration is avoided. This will require some pre-modeling work before the experiment.

**Acknowledgements**

Special thanks should be given to James MacLachlan of Fermilab, who used his expertise and devoted considerable time to help the authors to understand the problem. We are also grateful to Jim Norem of Argonne National Laboratory, without whose recommendations and help this work could not have been completed so expeditiously.

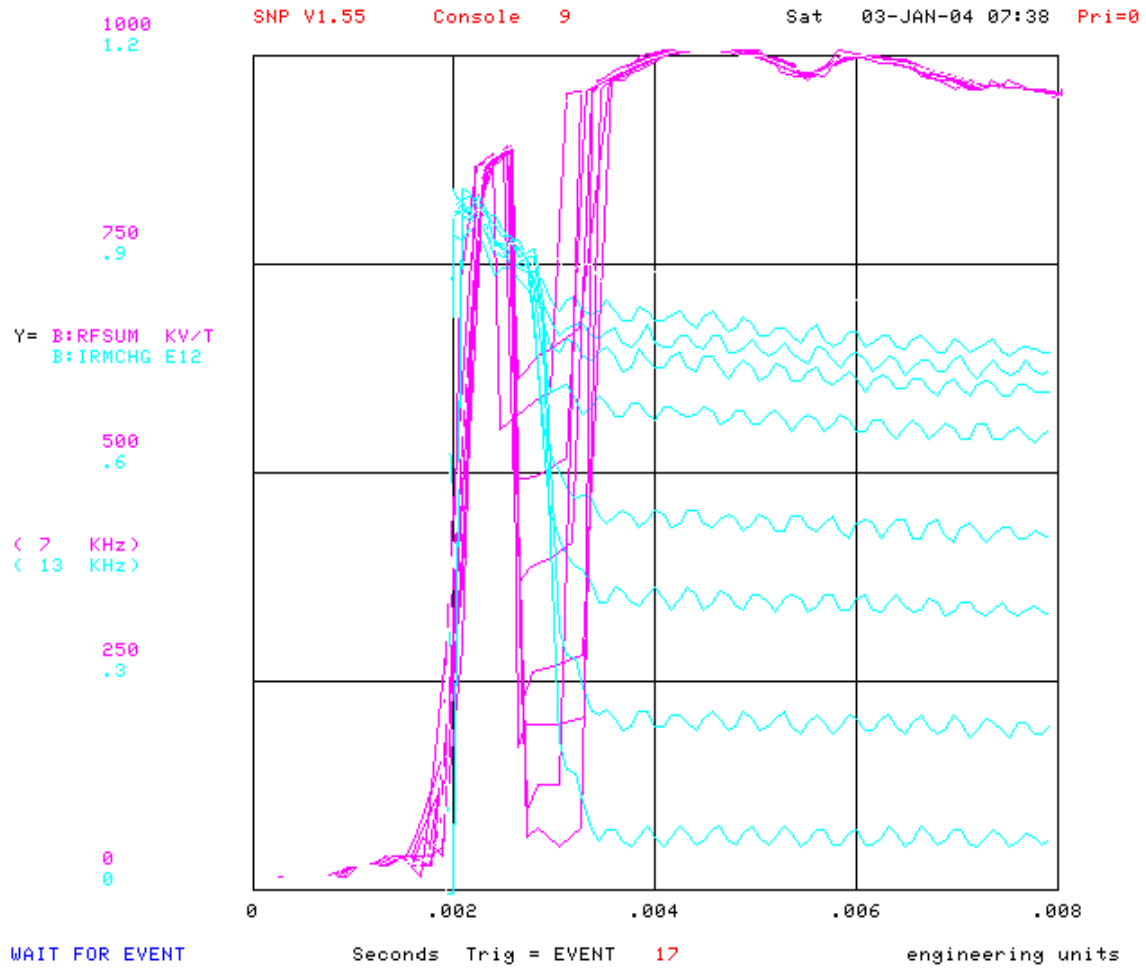
SNP V1.55 Console 9 Sat 03-JAN-04 07:38 Pri=0
Y= B:RFSUM KV/T
B:IRMCHG E12
Seconds Trig = EVENT 17
engineering units
WAIT FOR EVENT
1(a)

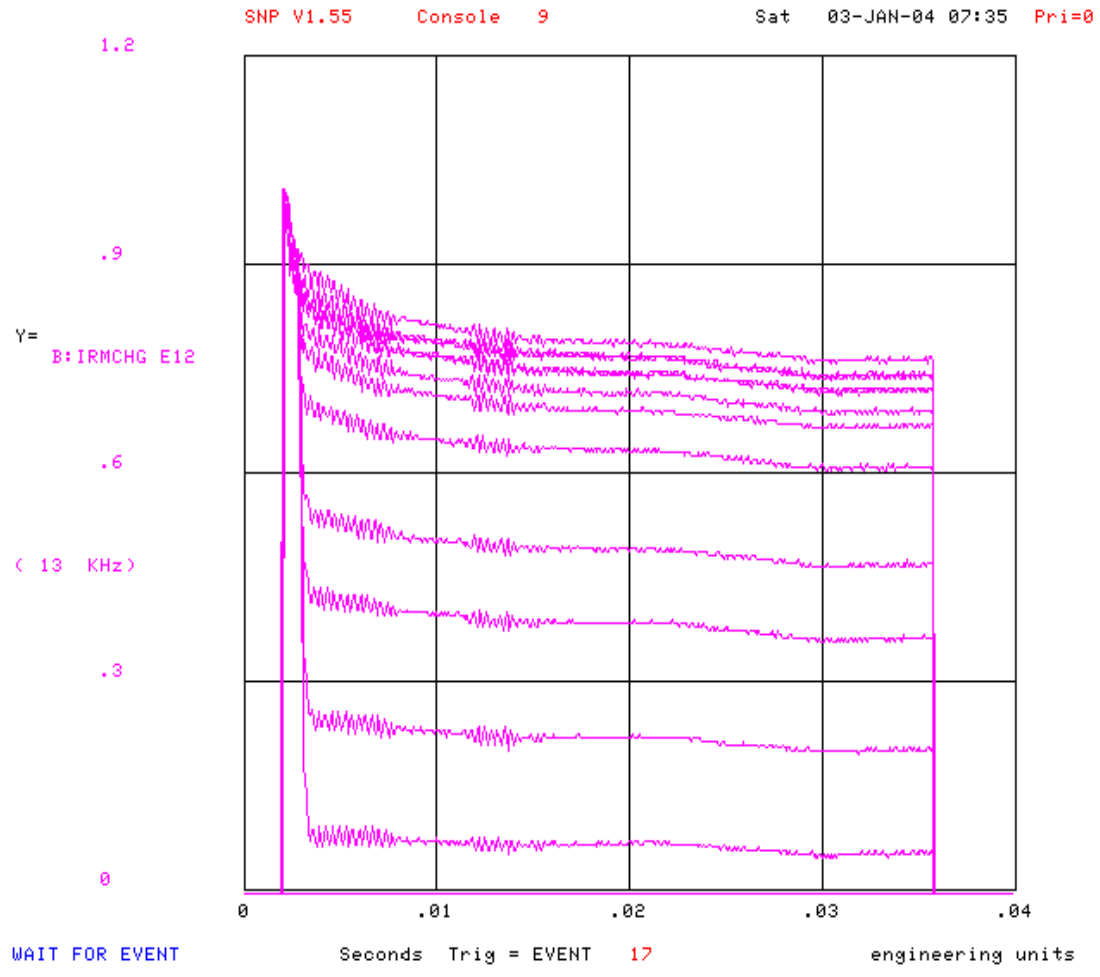
SNP V1.55 Console 9 Sat 03-JAN-04 07:35 Pri=0
Y= B:IRMCHG E12
Seconds Trig = EVENT 17
engineering units
WAIT FOR EVENT
1(b)

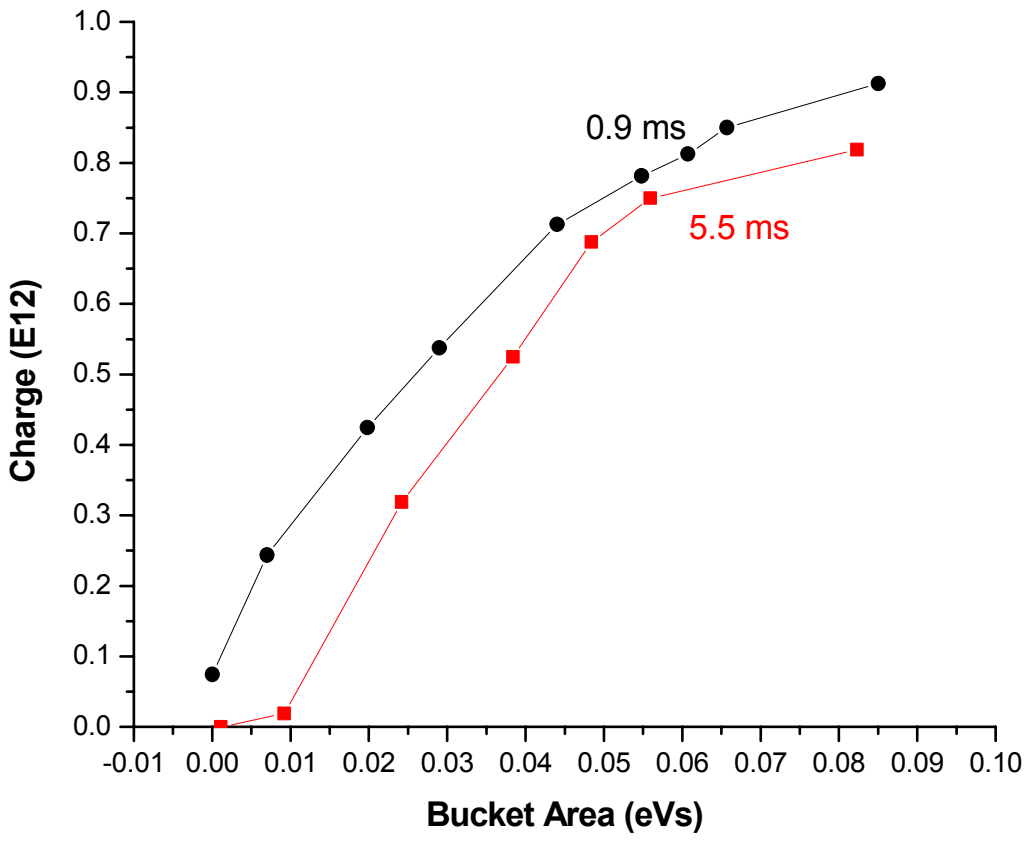
Charge (E12)
Bucket Area (eVs)
0.9 ms
5.5 ms
1(c)

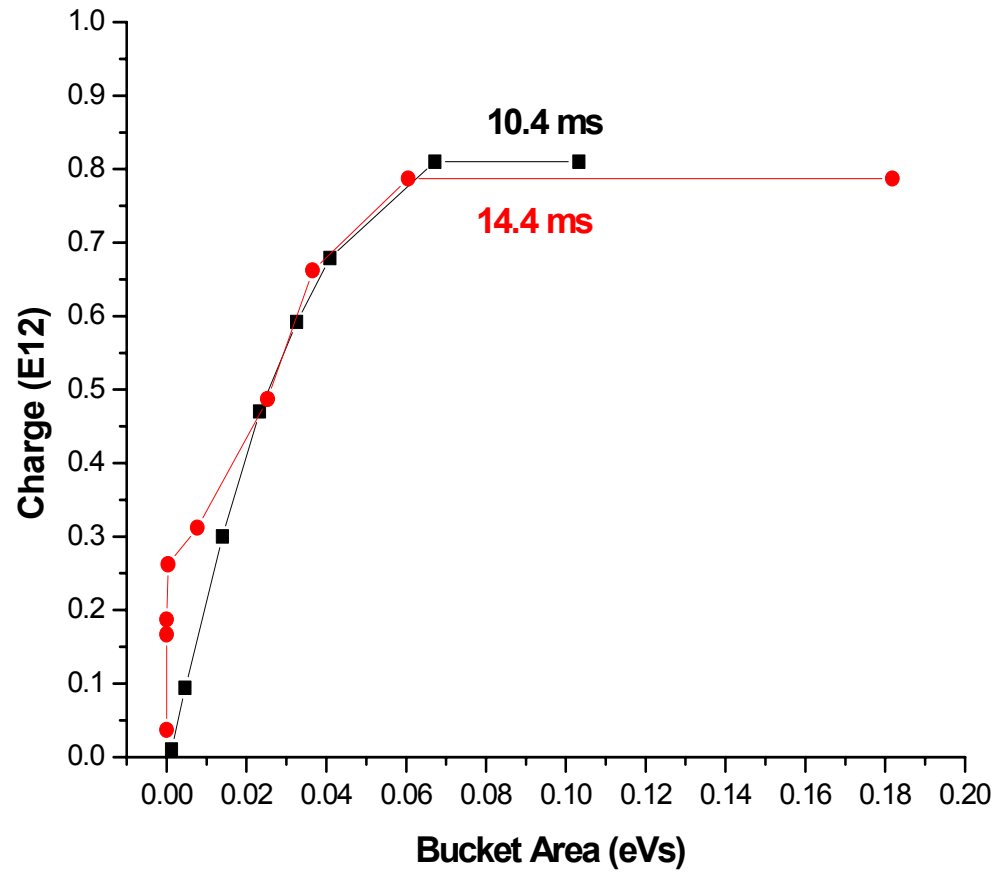
Charge (E12)
Bucket Area (eVs)
10.4 ms
14.4 ms
1(d)

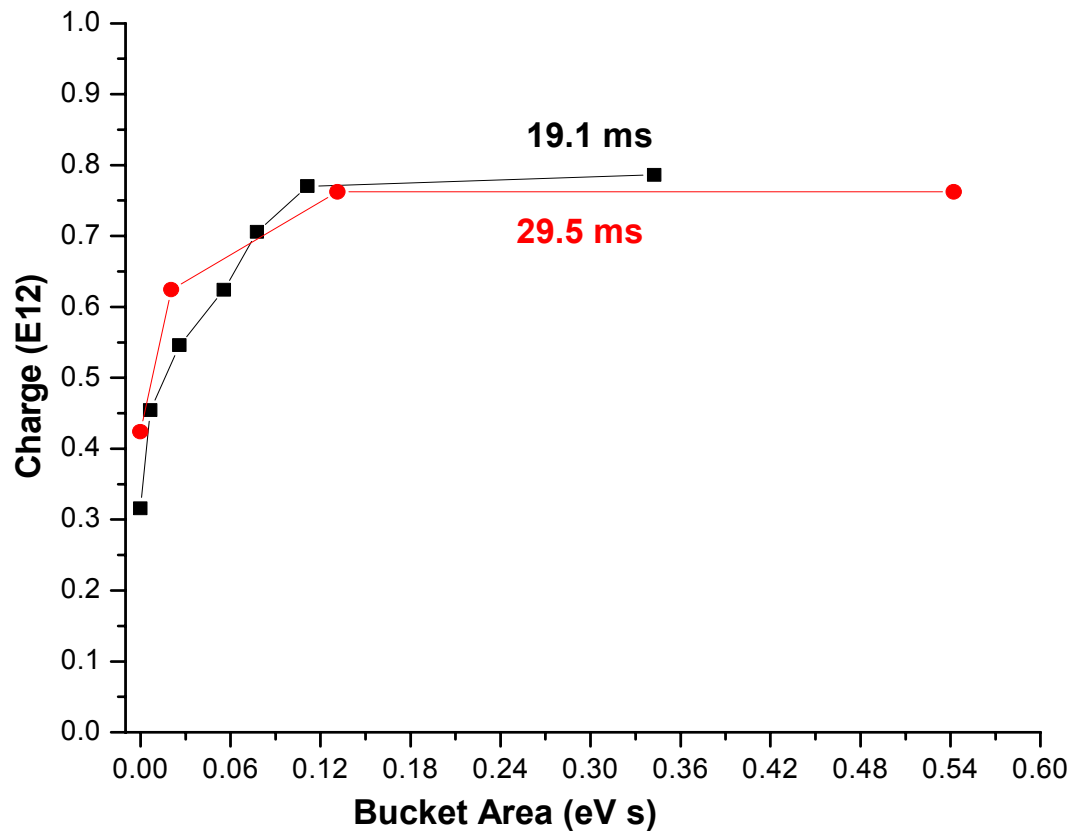
Charge (E12)
Bucket Area (eV s)
19.1 ms
29.5 ms
1(e)

Fig. 1(a) RFSUM was reduced to eight different values at about 0.9 ms after injection at low beam intensity with two turns injected. 1(b) Corresponding charge (i.e. beam intensity) signals were recorded for the RFSUM reduction at Fig. 1(a). 1(c) Charge *vs*. bucket area at 0.9 ms and 5.5 ms after injection. 1(d) Charge *vs*. bucket area at 10.4 ms and 14.4 ms after injection. 1(e) Charge *vs*. bucket area at 19.1 ms and 29.5 ms after injection.

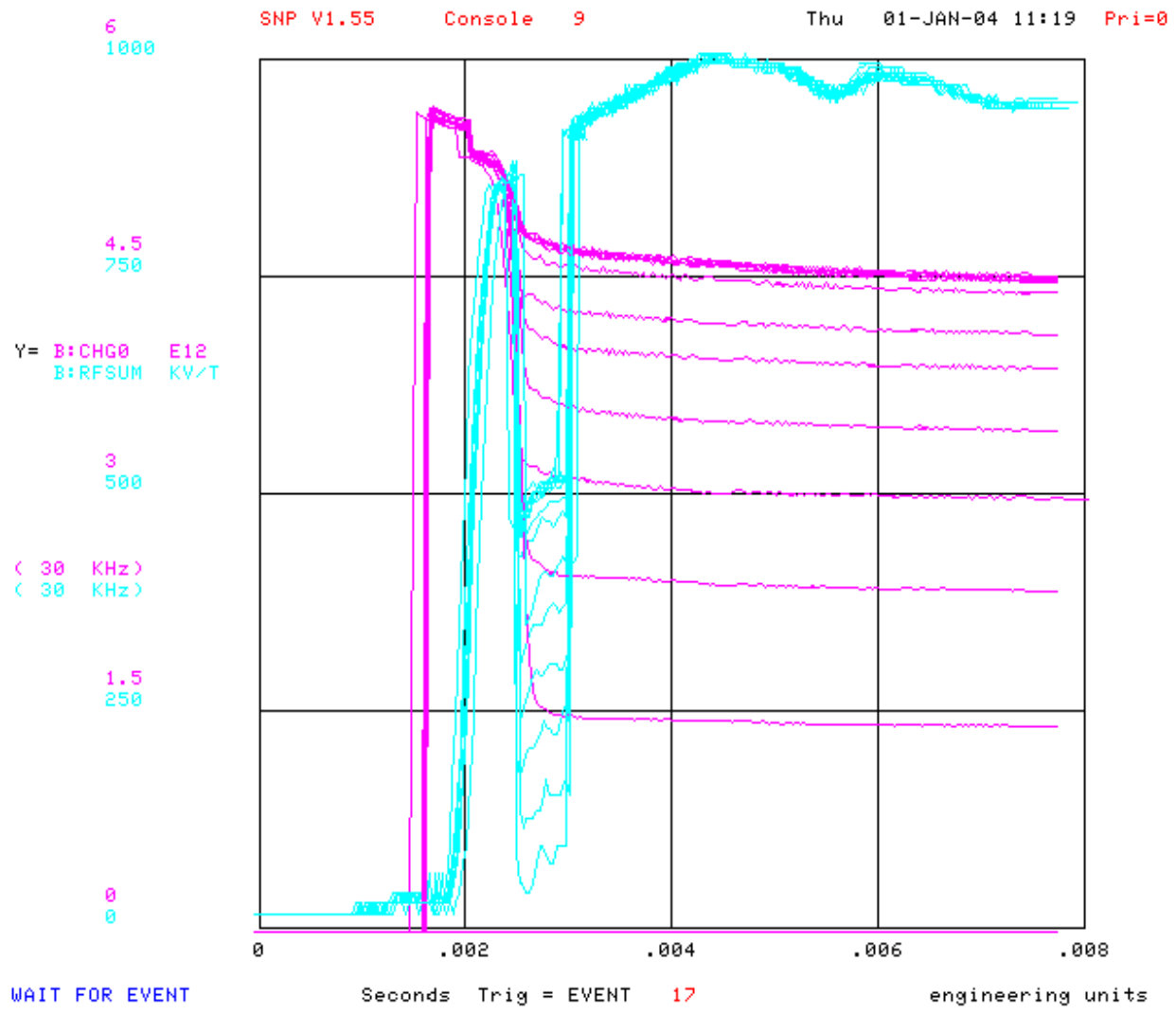

2(a)

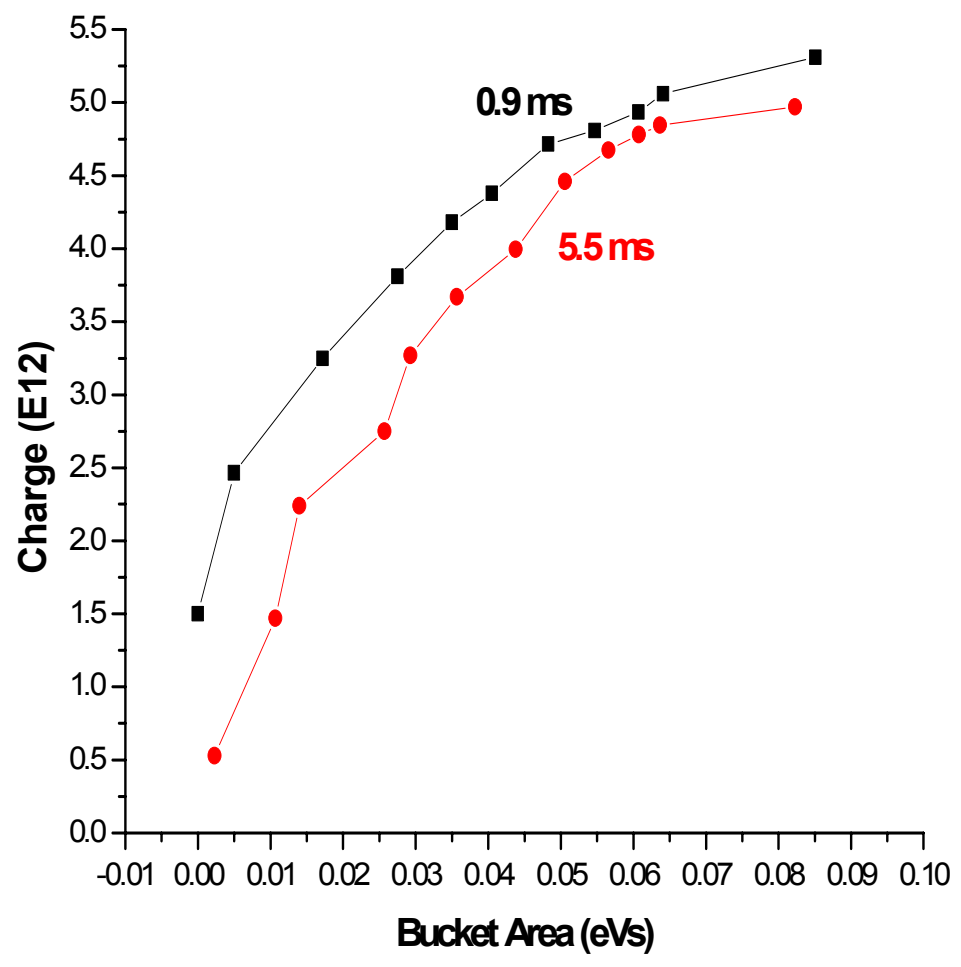

2(b)

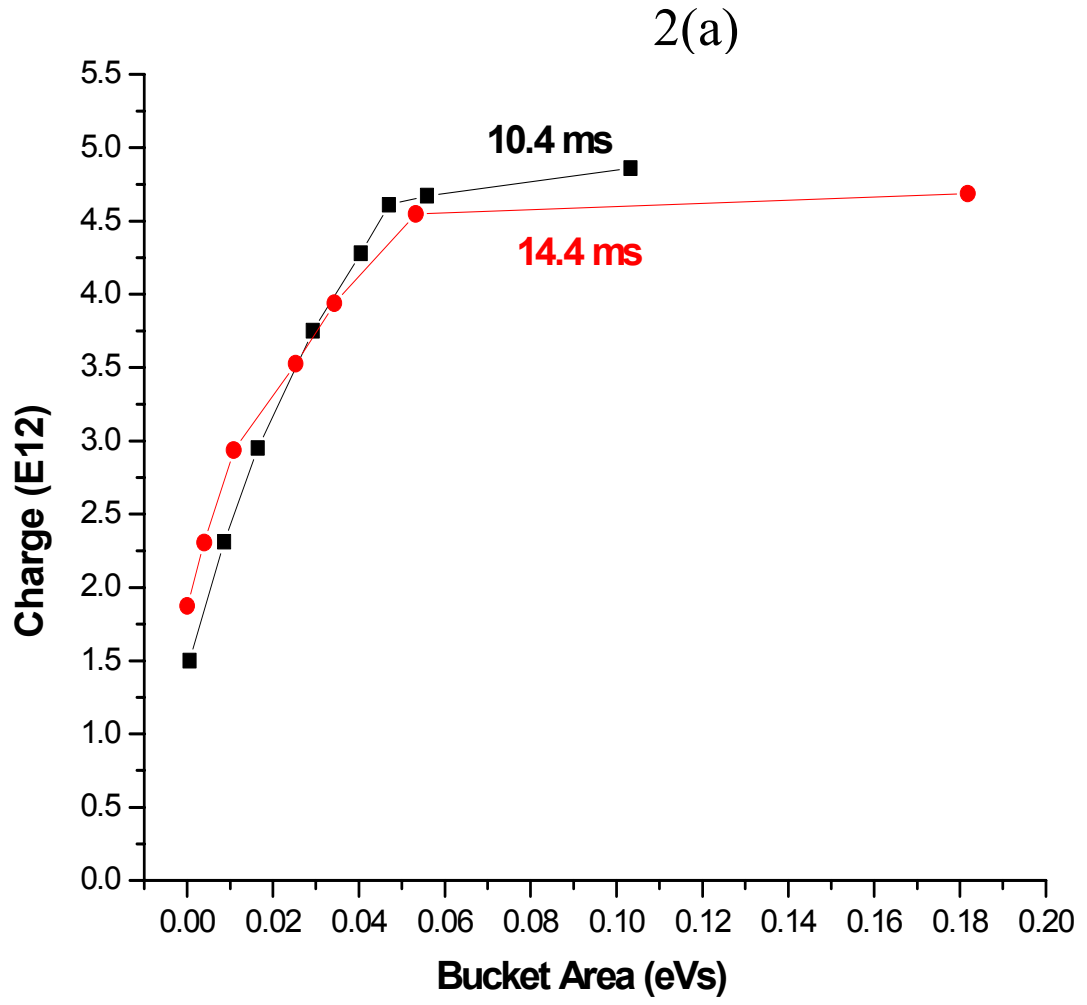

2(c)

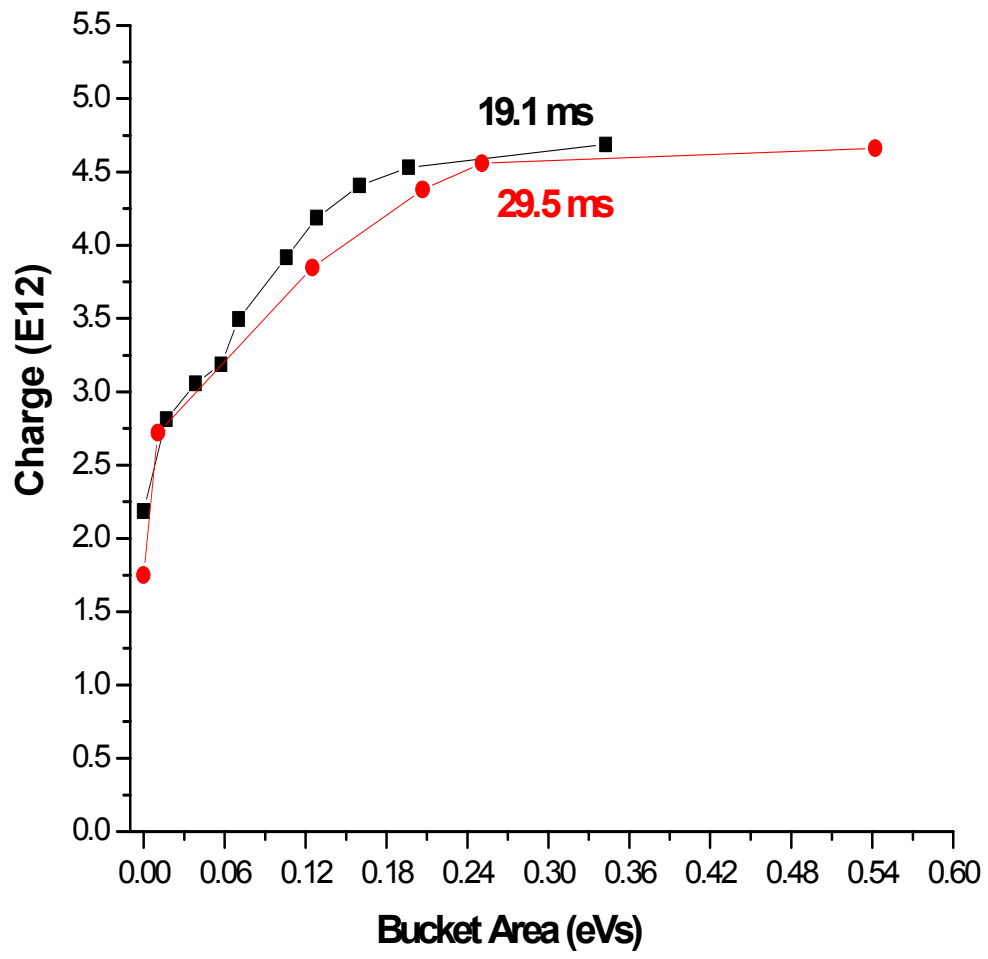

2(d)

Fig. 2(a) RFSUM was reduced to eight different values at about 0.9 ms after injection at high beam intensity with eleven turns injected. 2(b) Charge *vs*. bucket area at 0.9 ms and 5.5 ms after injection. 2(c) Charge *vs*. bucket area at 10.4 ms and 14.4 ms after injection. 2(d) Charge *vs*. bucket area at 19.1 ms and 29.5 ms after injection.

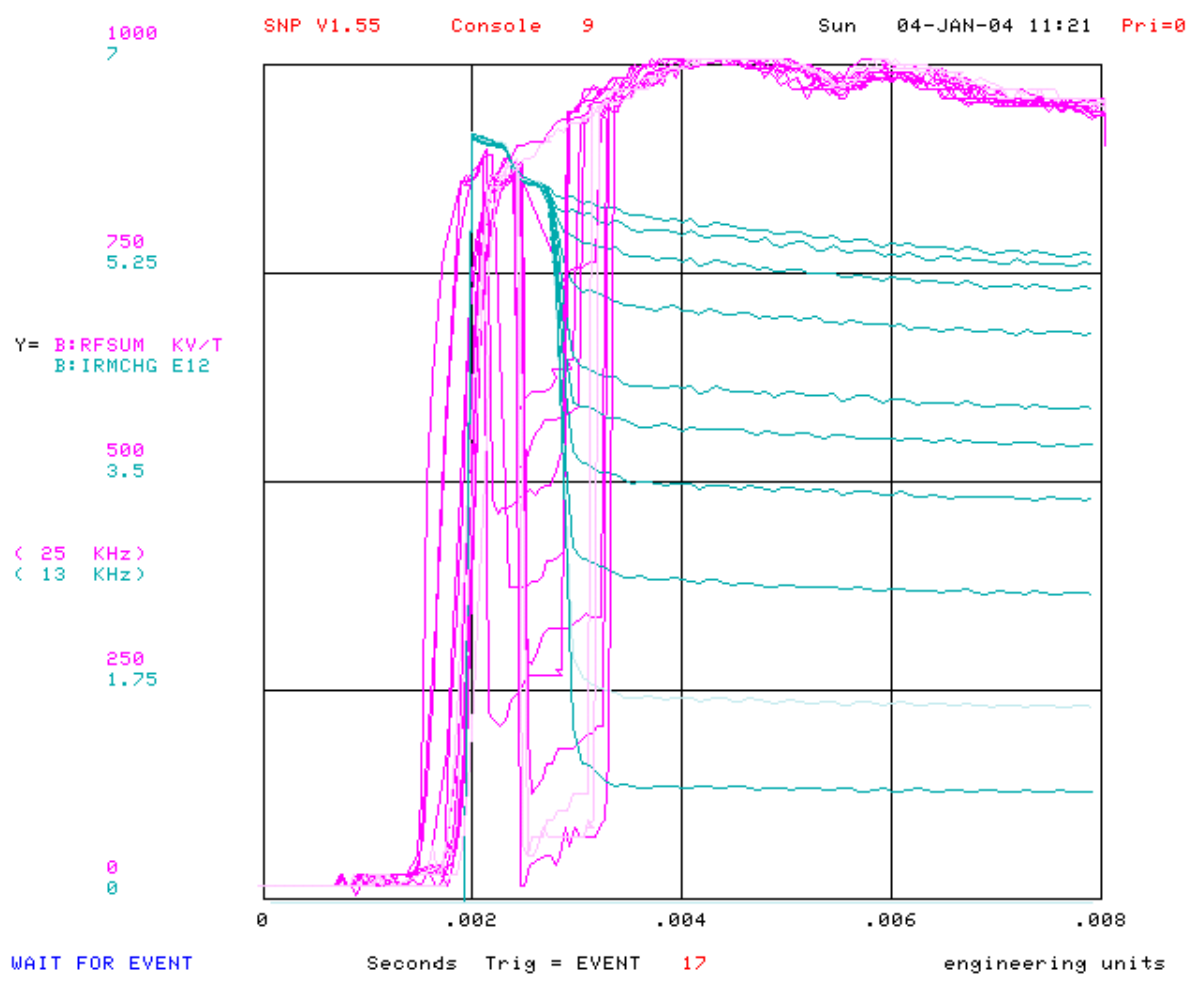

3(a)

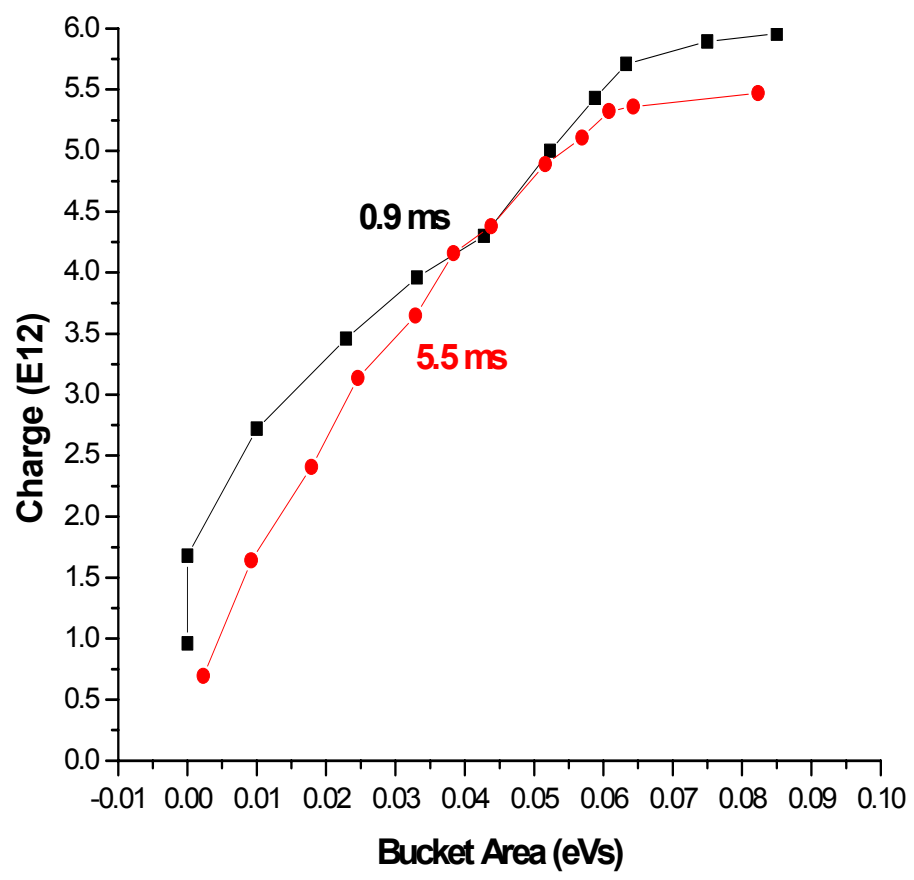

3(b)

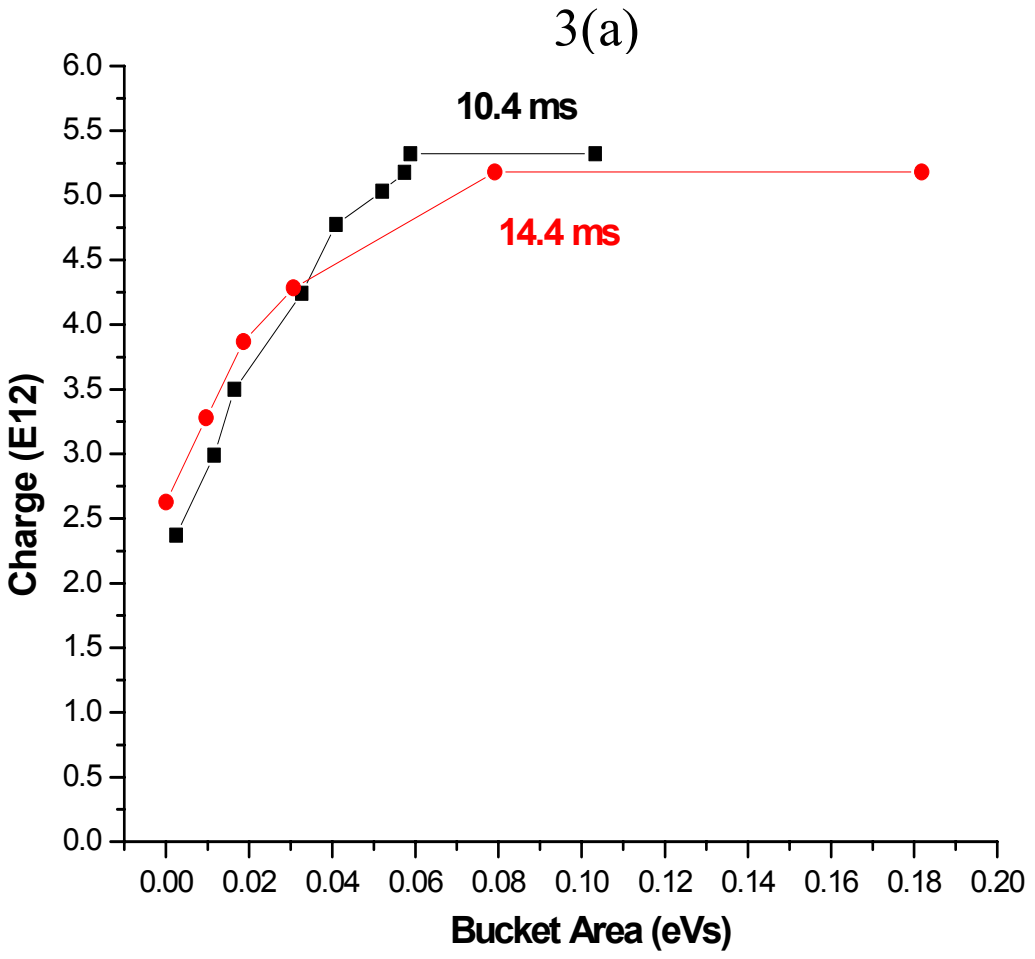

3(c)

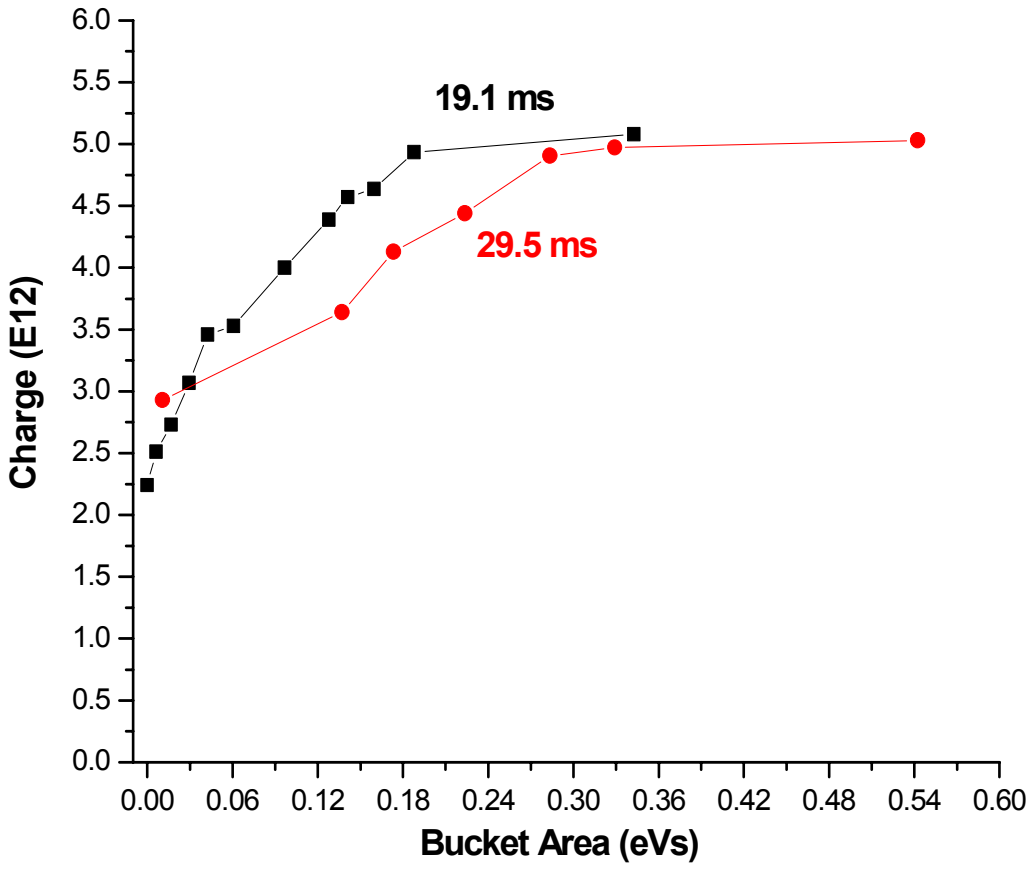

3(d)

Fig. 3(a) RFSUM was reduced to eight different values at about 0.9 ms after injection at high beam intensity with twelve turns injected. 3(b) Charge (i.e. beam intensity) *vs*. bucket area at 0.9 ms and 5.5 ms after injection. 3(c) Charge *vs*. bucket area at 10.4 ms and 14.4 ms after injection. 3(d) Charge *vs*. bucket area at 19.1 ms and 29.5 ms after injection.

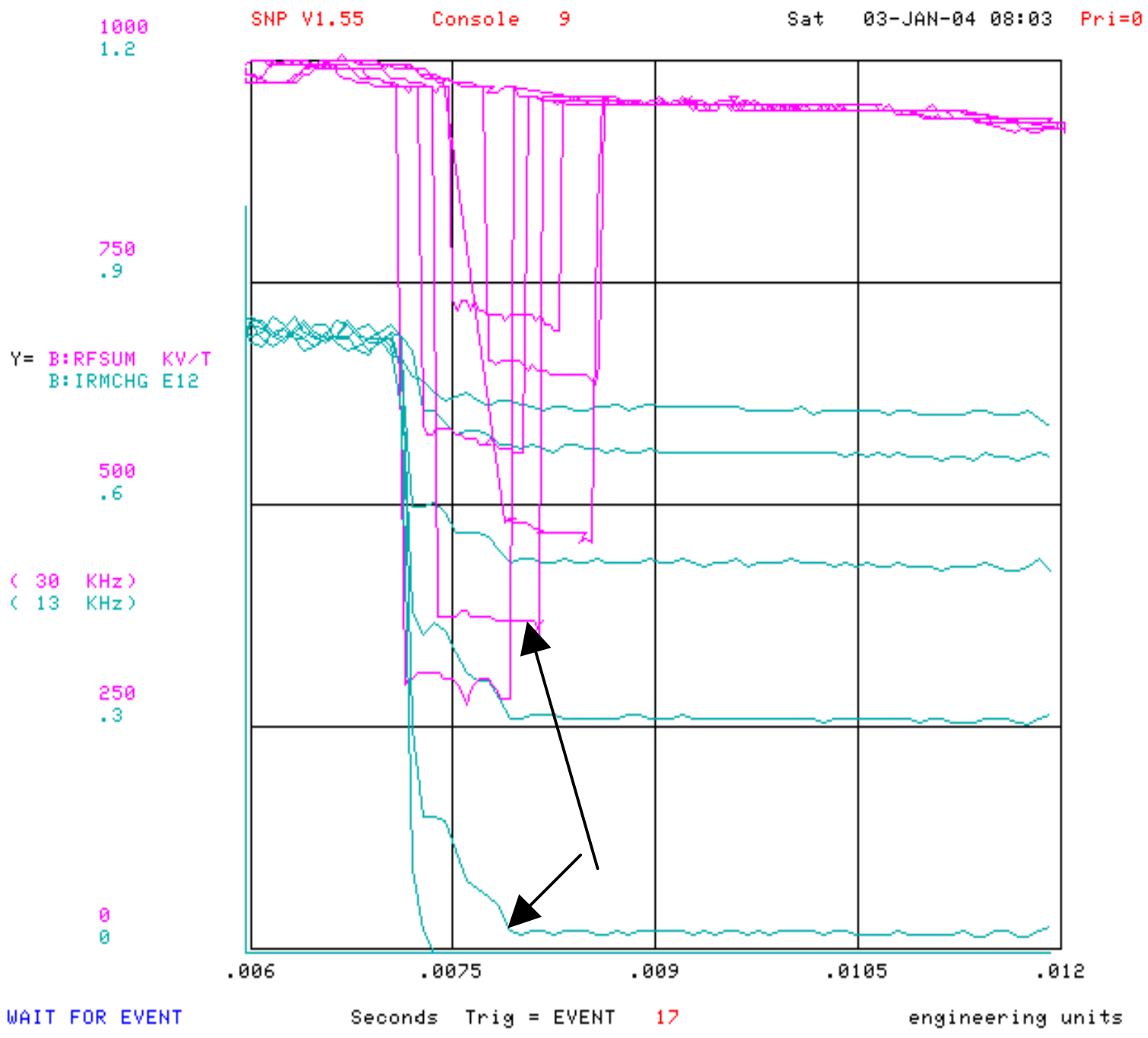

Fig. 4 Note the curves indicated by the two black arrows. When RFSUM was kept to the minimum, the continuous charge loss was caused by scraping.

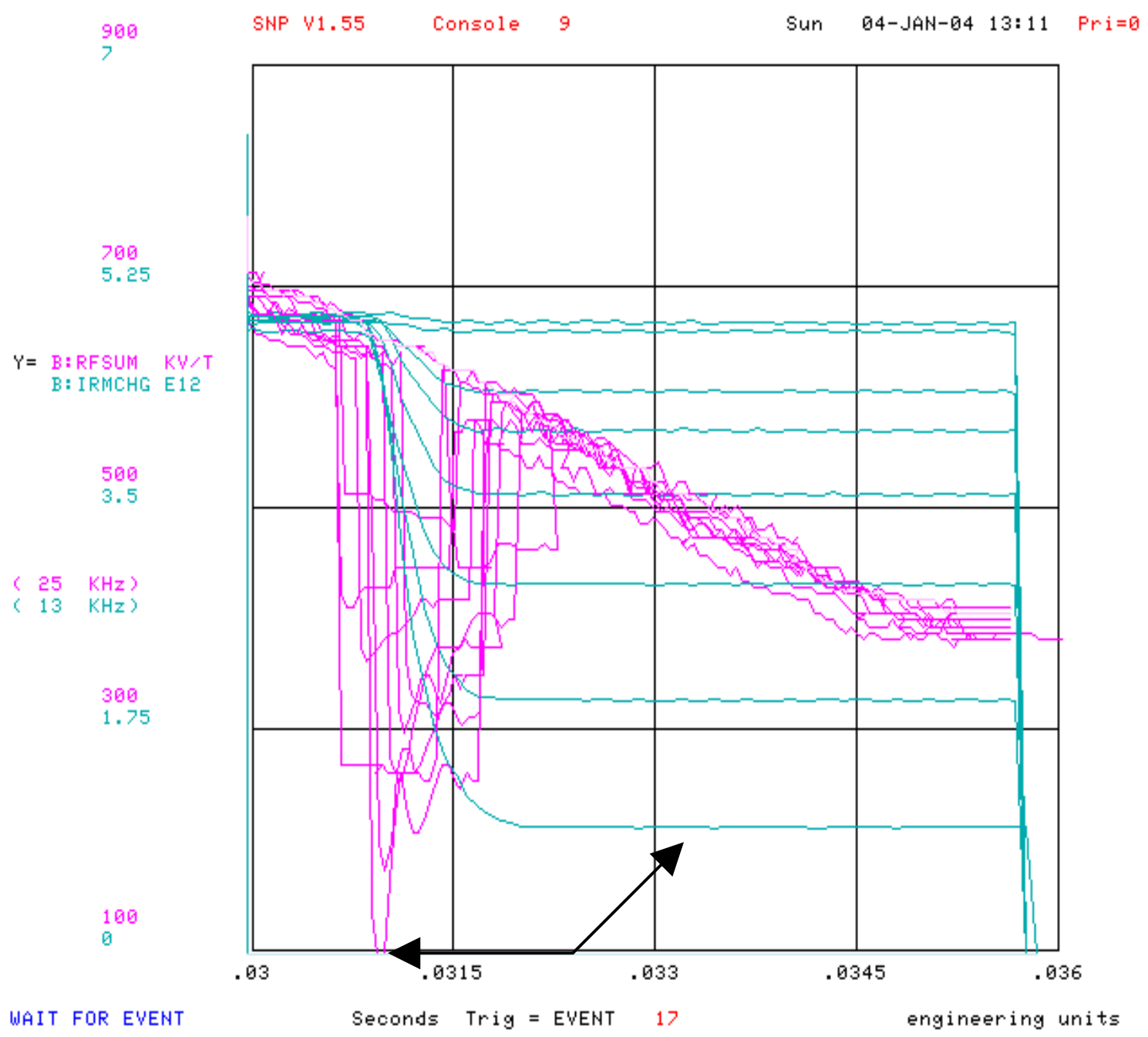

Fig. 5 Note the curves indicated by the two black arrows. There were still protons left after RFSUM was reduced to zero.